\documentclass[aps,superscriptaddress,twocolumn]{revtex4}

\usepackage{graphicx,epsfig}
\usepackage{color}
\usepackage[usenames,dvipsnames]{xcolor}
\usepackage{amsmath,bbm,amssymb, amsthm}
\usepackage{dsfont} 

\def\be{\begin{eqnarray}}
\def\ee{\end{eqnarray}}
\newcommand{\ket}[1]{| #1 \rangle}
\newcommand{\bra}[1]{\langle #1 |}

\newcommand{\idop}{\mathds{1}}

\begin{document}

\title{Extracting work from quantum measurement in Maxwell's demon engines}

\author{Cyril Elouard}
\affiliation{CNRS and Universit\'e Grenoble Alpes, Institut N\'eel, F-38042 Grenoble, France}

\author{David Herrera-Mart\'i}
\affiliation{CNRS and Universit\'e Grenoble Alpes, Institut N\'eel, F-38042 Grenoble, France}

\author{B. Huard}
\affiliation{Laboratoire de Physique, Ecole Normale Sup\'erieure de Lyon, 46 all\'ee d'Italie, 69364 Lyon Cedex 7, France}
\affiliation{Laboratoire Pierre Aigrain, Ecole Normale Sup\'erieure-PSL Research University, CNRS, Universit\'e Pierre et Marie Curie-Sorbonne Universit\'es, Universit\'e Paris Diderot-Sorbonne Paris Cit\'e, 24 rue Lhomond, 75231 Paris Cedex 05, France}

\author{Alexia Auff\`eves}
\email{alexia.auffeves@neel.cnrs.fr}
\affiliation{CNRS and Universit\'e Grenoble Alpes, Institut N\'eel, F-38042 Grenoble, France}

\date{\today}

\begin{abstract}
The essence of both classical and quantum engines is to extract useful energy (work) from stochastic energy sources, e.g. thermal baths. In Maxwell's demon engines, work extraction is assisted by a feedback control based on measurements performed by a demon, whose memory is erased at some nonzero energy cost. Here we propose a new type of quantum Maxwell's demon engine where work is directly extracted from the measurement channel, such that no heat bath is required. We show that in the Zeno regime of frequent measurements, memory erasure costs eventually vanish. Our findings provide a new paradigm to analyze quantum heat engines and work extraction in the quantum world.
\end{abstract}

\maketitle

\noindent {\it Introduction.} Thermodynamics was originally developed to optimize machines that would extract work from reservoirs at various temperatures, by exploiting the transformations of some working agent. These machines may be assisted by a so-called Maxwell's demon, that exploits information acquired on the agent to enhance work extraction, at the energy expense of resetting the demon's memory. Maxwell's demons and Szilard's engines have been investigated in several theoretical proposals \cite{Maxwell,Zurek-84,Marumaya-09,Mandal-12,Mandal-13,Deffner-13, Deffner-13bis, Strasberg-13, Horowitz-13, Barato-13}, including the thermodynamics of feedback control \cite{Sagawa-08,Sagawa-10,Sagawa-12}, and experimentally realized in various systems, e.g. Brownian particles~\cite{Toyabe-10,Roldan-14}, single electron transistors~\cite{Koski-14,Koski-15} and visible light~\cite{Vidrighin-16}. Latest experiments have started addressing the regime where the working agent exhibits quantum coherences \cite{Camati-16,Cottet-17}. The potential extraction of work from quantum coherence leads to interesting open questions related to the energetic aspects of quantum information technologies~\cite{Oppenheim-00, Del Rio-11, Park-13, Lostaglio-15, Kammerlander-16, Korzekwa-16}. Furthermore, novel designs for quantum engines, based on various kinds of quantum non-equilibrium reservoirs have been suggested~\cite{Scully-03,Abah-14,Zhang-14,Uzdin-15,Niedenzu-15,Manzano-16} and experimentally investigated \cite{Rossnagel-14, Rossnagel-16}.

\begin{figure}[h!]
\begin{center}
\includegraphics[width=8cm]{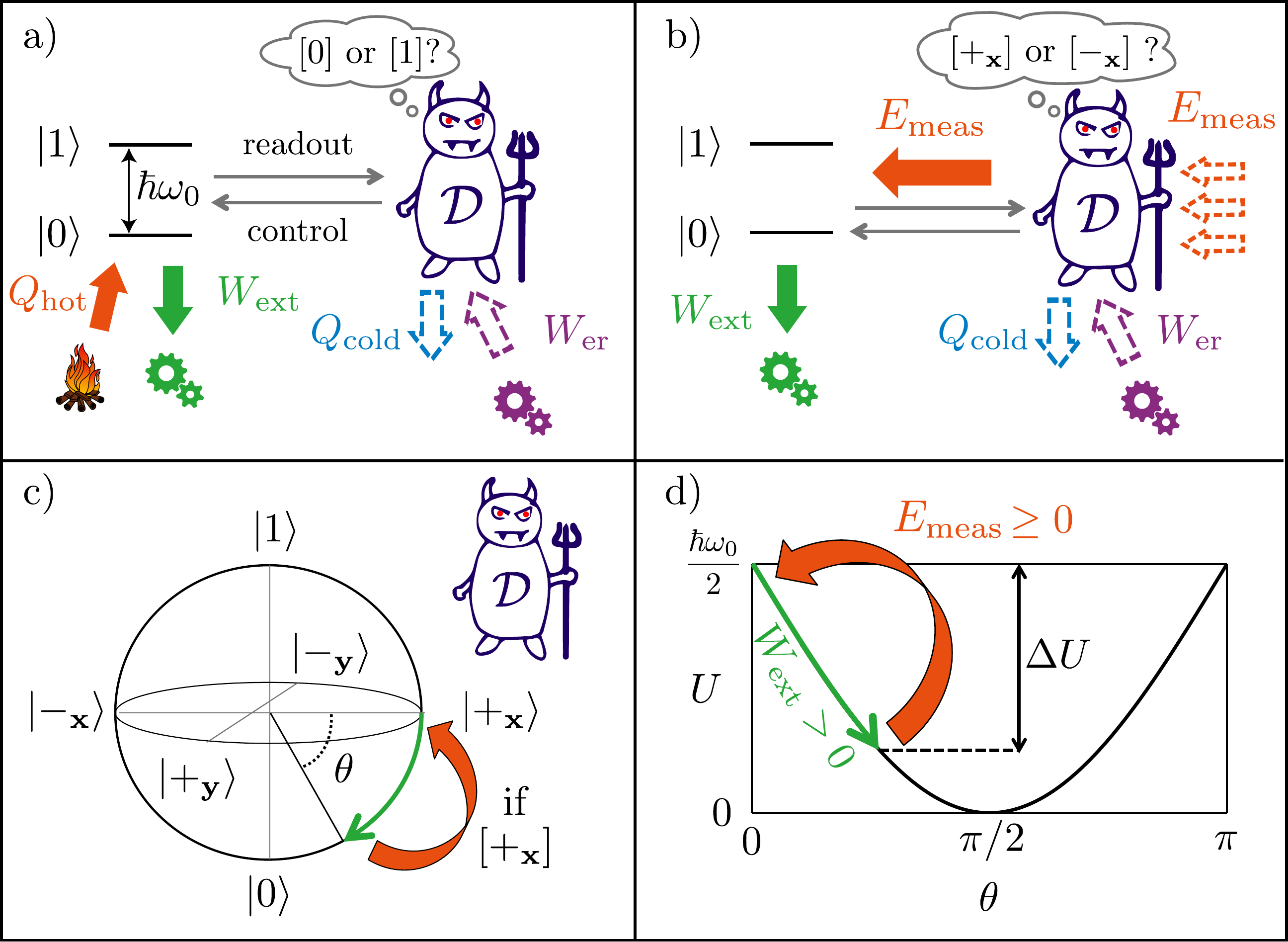}
\end{center}
\caption{Maxwell's demon assisted engines. a) Thermally driven engine. The working agent is a qubit of transition frequency $\omega_0$. A demon measuring in the qubit energy basis $\{ \ket{0}; \ket{1} \}$ allows to convert the heat $Q_\textrm{hot}$ extracted from a bath into work $W_\text{ext}$. The demon's memory is erased by some extra-work source at the minimal work cost $W_\text{er}$ while the heat $Q_\textrm{cold}$ is evacuated in a hidden cold bath. b) Measurement Powered Engine. In case the demon measures in the $\{\ket{+_\textbf{x}};\ket{-_\textbf{x}}\}$ basis, work can be extracted from the measurement channel and no hot bath is required. c) Evolution of the qubit state in the Bloch sphere. Between two measurements delayed by $\tau_w$, the evolution induced by the driving field is a rotation around the $Y$ axis at Rabi frequency $\Omega$ green thin arrow). The measurement projects the qubit on states $\ket{\pm_\textbf{x}}$ (orange thick arrow).  d) Evolution of the qubit internal energy $U(t)=\text{Tr}[\rho(t)H(t)]$ as a function of the Rabi angle $\theta = \Omega \tau_w$. During the Hamiltonian evolution the work $W_\text{ext} = - \Delta U$ is extracted into the driving field. The measurement provides the energy $E_\mathrm{meas}=W_\text{ext}$ back to the qubit. 
\label{fig1} }
\end{figure}

Most quantum engines considered so far involve a hot reservoir, which is the primary source of energy. In this framework, measurements performed by the demon are practical steps where information is extracted, without changing the energy of the working agent. Ultimately, measurement (just like decoherence) can appear as a detrimental step of the thermodynamic cycle as it destroys quantum coherences, further preventing to extract work from them \cite{Lloyd-97}. Here we adopt a different approach and show that measurement itself can be exploited as a fuel in a new kind of quantum engine. Originally here, the demon can perform measurements that are sensitive to states in an arbitrary basis of the system Hilbert space. It was recently shown~\cite{Brandner-15, Elouard-16,Abdelkhalek17,Yi17} that performing projective measurements on a quantum system can change its average energy, provided that the measured observable does not commute with the system Hamiltonian. If the demon projects the system state onto superpositions of energy eigenstates, it can thus provide energy just by measuring. We study the performance of the engine as a function of the measurement basis and repetition rate, and evidence that a net work can be extracted from the measurement channel, even in the absence of any hot reservoir that would be directly coupled to the qubit. In the Zeno limit of quickly repeating measurements, the entropy of the demon's memory vanishes, suppressing the energetic costs related to its erasure.  \\

\noindent {\it Thermally driven engine} Before detailing our proposal, we recall a possible operating mode for an elementary thermally driven engine assisted by a Maxwell's demon (Fig.~\ref{fig1}a). The working agent is a qubit whose energy eigenstates are $\ket{0}$ and $\ket{1}$, and transition frequency is $\omega_0$. The bare system Hamiltonian is $H_0 = \hbar \omega_0\ket{1}\bra{1}$. Before the cycle starts, the qubit is coupled to a hot bath at temperature $T_\text{hot}$ verifying $k_\text{B} T_\text{hot}\gg \hbar \omega_0$ where $k_\text{B}$ is the Boltzmann constant. It is thus thermalized in the mixed state $\rho_\text{q}(0)=\idop/2$ where $\rho_\text{q}(t)$ is the qubit density matrix. During this heating step the qubit entropy $S_\text{q}(t)=-k_\text{B}\text{Tr}[\rho_\text{q}(t) \log(\rho_\text{q}(t))]$ (resp. mean internal energy $U(t) = \text{Tr}[\rho(t)H_0]$) increases up to $S_\text{q}(0)=k_\text{B}\log(2)$ (resp. $U(0)=\hbar \omega_0/2=Q_\text{hot}$, where $Q_\text{hot}$ is the heat extracted from the hot bath). 

The qubit is then decoupled from the bath and measured by a demon ${\cal D}$ in its energy basis. We first do not focus on the physical implementation of the demon and treat it as a device extracting and storing information on the qubit onto some classical memory (readout), and exerting some action on the qubit, conditioned to the readout (control). The physical durations of the readout, feedback and erasure steps are neglected, and additional energetic costs related to amplification of measurement outcomes are not considered. After the readout step, the demon's memory is perfectly correlated with the qubit state, and its entropy $S_{\cal D}$ satisfies $S_{\cal D}=S_\text{q}(0)=k_\text{B}\log(2)$. 

The work extraction step is triggered if the qubit is measured in the state $\ket{1}$, which happens half of the times since we have considered large temperatures $T_\text{hot}$. A convenient way to extract work consists in resonantly coupling the qubit to a classical drive $H_\text{c}(t) = i(\hbar \Omega/2) (\sigma_- e^{i\omega_0 t} - \sigma^\dagger_- e^{-i\omega_0 t})$, where $\sigma_-= \ket{0}\bra{1}$ is the qubit lowering operator and $\Omega$ the Rabi frequency. In the frame rotating at the drive frequency $\omega_0$, the Hamiltonian eigenstates are $\ket{\pm_\textbf{y}}$. We have defined $\ket{+_\textbf{n}}= e^{-i\phi_\textbf{n}/2} \cos(\theta_\textbf{n}/2) \ket{1} + e^{i\phi_\textbf{n}/2} \sin(\theta_\textbf{n}/2) \ket{0}$, with the normalized vector $\textbf{n} = \left( \sin(\theta_\textbf{n}) \cos(\phi_\textbf{n}), \sin(\theta_\textbf{n}) \sin(\phi_\textbf{n}), \cos(\theta_\textbf{n}) \right)$ being written in the $(x,y,z)$ basis of the Bloch sphere, and $\textbf{y} = (0,1,0)$. The coupling time $\tau_{\pi}$ is tuned such that the qubit undergoes a $\pi$-pulse that coherently brings the system from state $\ket{1}$ to $\ket{0}$.

The change of internal qubit energy, induced by the $\pi$ pulse, reads $U(\tau_{\pi}^+)-U(\tau_{\pi}^-) = -\hbar \omega_0$ if the qubit is initially measured in $\ket{1}$ (zero otherwise). This energy decrease quantifies the extracted work, that is used to coherently amplify the driving field of the $\pi$ pulse by one extra photon. In practice, the qubit could then provide work to power up another processing unit of a quantum machine in photonic or microwave circuits~\cite{Ikonen-16}. Note that we consider large enough drive amplitudes, so that the extra photon has negligible effect on the coupling Hamiltonian. The mean extracted work finally equals $W_\text{ext} =\hbar \omega_0/2 = Q_\text{hot}$. The cycle is closed with the erasure of the demon's classical memory. This step can be realized by performing Landauer's protocol \cite{Landauer, Berut-12,Del Rio-11}, which requires some hidden work source and cold bath of temperature $T_{\cal D}$ as additional resources. When it is performed quasi-statically, erasure costs a minimal work $W_\text{er}=-Q_\text{cold}=S_{\cal D} T_{\cal D}$. %Note that this work is of different nature as the extracted work.

Finally, we define the engine's yield $\eta_\text{cl}$ {\color{black} as the difference between the extracted work $W_\text{ext}$ and the work required to erase the memory $W_\text{er}$, divided by the resource, i.e. the heat $Q_\text{hot}$ \cite{Lloyd-97}. We find
\begin{equation}
\eta_\text{cl} = 1- \frac{2 k_\text{B} T_{\cal D} \log(2)}{\hbar \omega_0}. \label{classicaleff}
\end{equation}
}
\noindent  %As for Carnot engines, a maximum value of $\eta_\text{cl}=1$ can be reached in the limit $T_{\cal D}\to 0$. However the condition of reversibility would then impose a vanishing engine power, since the duration of erasure diverges.
\\

\noindent {\it Measurement powered engine (MPE)} We now introduce a new protocol to operate the engine, for which the demon is allowed to perform projective measurements in arbitrary bases $\{\ket{+_\textbf{n}};\ket{-_\textbf{n}} \}$ of the system state. The measurement gives rise to some energy change $E_\text{meas}$ \cite{Brandner-15, Elouard-16,Abdelkhalek17,Yi17}: This is the only fuel for our engine, and no hot thermal bath is required (Fig.~\ref{fig1}b). We first focus on the case $\textbf{n} = \textbf{x} = (1,0,0)$ which will reveal to maximize the efficiency. The qubit is initially measured and prepared in the operating point $\ket{+_\textbf{x}}$. A cycle then consists in the four following steps: 

\noindent (i){\it Work extraction} The qubit is coupled to the drive (Hamiltonian $H_\text{c}(t)$) during a time $\tau_\text{w}$. Introducing the Rabi angle $\theta = \Omega \tau_\text{w}$ (Fig.~\ref{fig1}c), the extracted work reads $W_\text{ext} = - U(\tau_\text{w})+U(0) =  \hbar \omega_0 \sin(\theta)/2$. $W_\text{ext}$ is strictly positive as long as $\theta \leq \pi$ (Fig.\ref{fig1}d). Each cycle provides the same amount of work, which is extracted from the coherence of the $\ket{+_\textbf{x}}$ state \cite{Cottet-17,Uzdin-15}. Reciprocally, starting from the state $\ket{-_\textbf{x}}$ would trigger energy absorption from the drive and negative work extraction. 

\noindent (ii){\it Readout}. The demon measures the qubit in the basis $\{\ket{+_\textbf{x}};\ket{-_\textbf{x}} \}$, preparing it in the mixed state $\rho_\text{q} = \cos^2(\theta/2)\ket{+_\textbf{x}}\bra{+_\textbf{x}} + \sin^2(\theta/2)\ket{-_\textbf{x}}\bra{-_\textbf{x}}$. The qubit states $\ket{+_\textbf{x}}$ and $\ket{-_\textbf{x}}$ are classically correlated with the states of the demon's memory: Therefore the entropies of the qubit and the demon satisfy $S_\text{q} = S_{\cal D} = k_\text{B} \log(2) H_2[\cos^2(\theta/2)]$, where $H_2[x] = -x\log_2(x)-(1-x)\log_2(1-x)$ is the Shannon entropy (expressed in bits). On the other hand, whatever its outcome, the measurement deterministically restores the qubit internal energy to its initial value since $U_{\pm_\textbf{x}} = \bra{\pm_\textbf{x}} H_0 \ket{\pm_\textbf{x}} = \hbar \omega_0/2$ , providing an energy $E_\text{meas} =-W_\text{ext}$. 

Importantly here, measurement plays three roles. First, as in classical Maxwell's demons engines, it allows extracting information on the qubit state. As a property of quantum measurement, it also increases the qubit entropy. Finally, it provides energy to the qubit since the measurement basis does not commute with the bare energy basis. These two last characteristics make the connection between the measurement process and the action of a thermal bath, which lies at the basis of our MPE.

\noindent(iii){\it Feedback}. If the outcome is $[-_\textbf{x}]$, a feedback pulse prepares the qubit back in state $\ket{+_\textbf{x}}$. This step has no energy cost, e.g. it can be realized by letting the qubit freely evolve (Rotation around the $Z$ axis of the Bloch sphere) during some appropriate time. At the end of this step the qubit is prepared back in the pure state $\ket{+_\textbf{x}}$.

\noindent(iv){\it Erasure}. The classical demon's memory is finally erased to close the cycle. Just like in Eq.~(\ref{classicaleff}), we consider the minimal bound for the erasure work $W_\text{er} = T_{\cal D} S_{\cal D}$, which is reached in quasi-static processes.

By applying the same definition as above with $E_\text{meas}$ as the resource, we find for the yield:

\begin{equation}
\eta(\theta) = 1-\frac{ 2 k_\text{B} T_{\cal D}\log(2)}{\hbar \omega_0}\frac{H_2[\cos^2(\theta/2)] }{ \sin(\theta)}.
\end{equation}

\begin{figure}[t]
\begin{center}
\includegraphics[width=8.2cm]{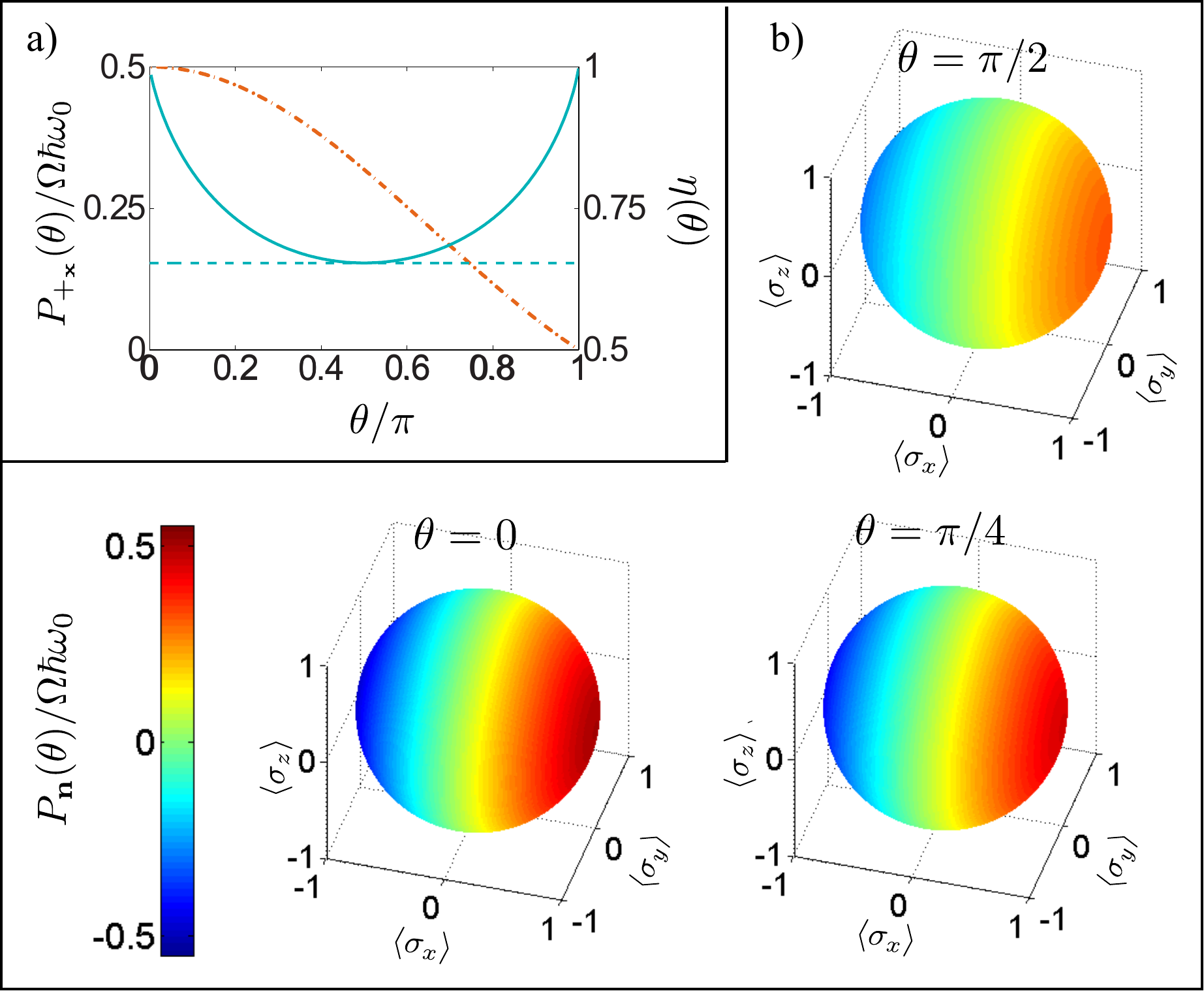}
\end{center}
\caption{Performance of the measurement powered engine. a) Orange dotted-dashed line: normalized extracted power $P_{+_\textbf{x}}/\Omega\hbar\omega_0$; Solid blue line:  MPE yield $\eta(\theta)$ as a function of the Rabi angle $\theta$; Dashed blue line:  thermally driven engine's yield $\eta_\text{cl}$. We chose $\hbar \omega_0/(2k_\text{B}T_{\cal D} \log(2))=2$. b) Normalized extracted power $P_{\textbf{n}}(\theta)/\Omega\hbar\omega_0$ in the Zeno limit where $\theta\rightarrow 0$, as a function of the chosen operating point $\ket{+_\textbf{n}}$ for $\theta = \pi/2$, $\theta = \pi/4$ and in the Zeno limit $\theta\to 0$.  \label{fig2} }
\end{figure}

\noindent The yield $\eta(\theta)$ is plotted in Fig.~\ref{fig2}a, together with the classical yield $\eta_\text{cl}$. It is minimized when $\theta=\pi/2$ where $\eta=\eta_\text{cl}$. There both the thermally driven engine and the MPE provide the same amount of mean extracted work $\hbar\omega_0/2$ and require the same work $W_\text{er} = k_\text{B} T_{\cal D} \log(2)$ to erase the demon's memory. When $\theta \neq \pi/2$, erasure cost decreases as the entropy of the memory decreases, leading to a larger yield than in the thermal case.

The limit $\theta\rightarrow 0$ corresponds to the Zeno regime where stroboscopic readouts are performed at a rate faster than the Rabi frequency ($\tau_\text{w} \ll \Omega^{-1}$). Here the extracted work per cycle scales as $\theta$, while the qubit is frozen in the $\ket{+_\textbf{x}}$ state and the energetic erasure cost behaves as $o(\theta)$, which maximizes the yield $\eta \rightarrow 1$. In this regime, the device behaves as a transducer converting deterministically the energy provided by the measurement channel into work. This defines another operating mode of the MPE, which has no classical equivalent as the control over the energy transfer is ensured by performing frequent quantum measurements. Interestingly, feedback is still crucial to allow work extraction in the steady state regime. Without feedback indeed, the energetic costs related to Zeno stabilization diverge \cite{Abdelkhalek17}. Reciprocally, it is possible to extract some positive work without feedback during some finite time before the engine switches off. An experimental realization of this case is studied below.

\noindent We finally study the extracted power $P_{+_\textbf{n}}(\theta)$ (See Supplemental and Fig.\ref{fig2}), as a function of the operating point $\ket{+_\textbf{n}}$ and Rabi angle $\theta$. As expected, measuring the qubit in its bare energy basis $\ket{\pm_\textbf{z}}$ leads to zero power extraction since the measurement channel does not provide any energy. The engine also switches off if the demon measures the qubit in the coupling Hamiltonian eigenbasis $\ket{\pm_\textbf{y}}$, since these states do not give rise to any work exchange between the qubit and the drive. As mentioned previously, maximal power is obtained if $\ket{+_\textbf{n}} = \ket{+_\textbf{x}}$. Here the qubit coherently provides energy to the drive in the fastest way. Reciprocally, using state $\ket{-_\textbf{x}}$ triggers the reverse mode where the engine coherently extracts maximal power from the drive.  \\

\noindent {\it Implementation} In the Supplemental we propose an experiment where work is extracted in the quantized field of a cavity mode. Here we consider a superconducting circuit where work is extracted as propagating photons \cite{Cottet-17}. A superconducting qubit is dispersively coupled to a single mode of a microwave cavity, such that 
by measuring the cavity state, one performs a Quantum Non Demolition projective measurement of the qubit $\sigma_Z$ observable \cite{Wallraff-04}. It is then possible to realize a periodic measurement of $\sigma_X$, and control feedback as demonstrated in Ref.~\cite{Campagne-13}. To do so, two short $\pi/2$ pulses are applied before and after the projective measurement of $\sigma_Z$. The first pulse coherently maps the targeted measurement basis $\{ \ket{+_\textbf{x}}; \ket{-_\textbf{x}} \}$ on the $\{ \ket{0}; \ket{1} \}$ basis, while the second pulse performs the reverse operation. The net effect of the total sequence is to project the qubit state onto one of the two eigenstates of $\sigma_X$. For optimized geometries~\cite{Walter-17} or using longitudinal coupling~\cite{Kerman,Billangeon,Didier-15,Richer}, Quantum Non Demolition projective measurements of $\sigma_Z$ can be performed and repeated on typical times as fast as $50$ ns, while $\pi/2$ pulses take of the order of 10~ns. Taking into account the physical durations of these different steps, the effective projective readout of $\sigma_X$ can be completed within a realistic time $\tau_\text{mes} = 70$~ns. 

\begin{figure}
\begin{center}
\includegraphics[width=8.2cm]{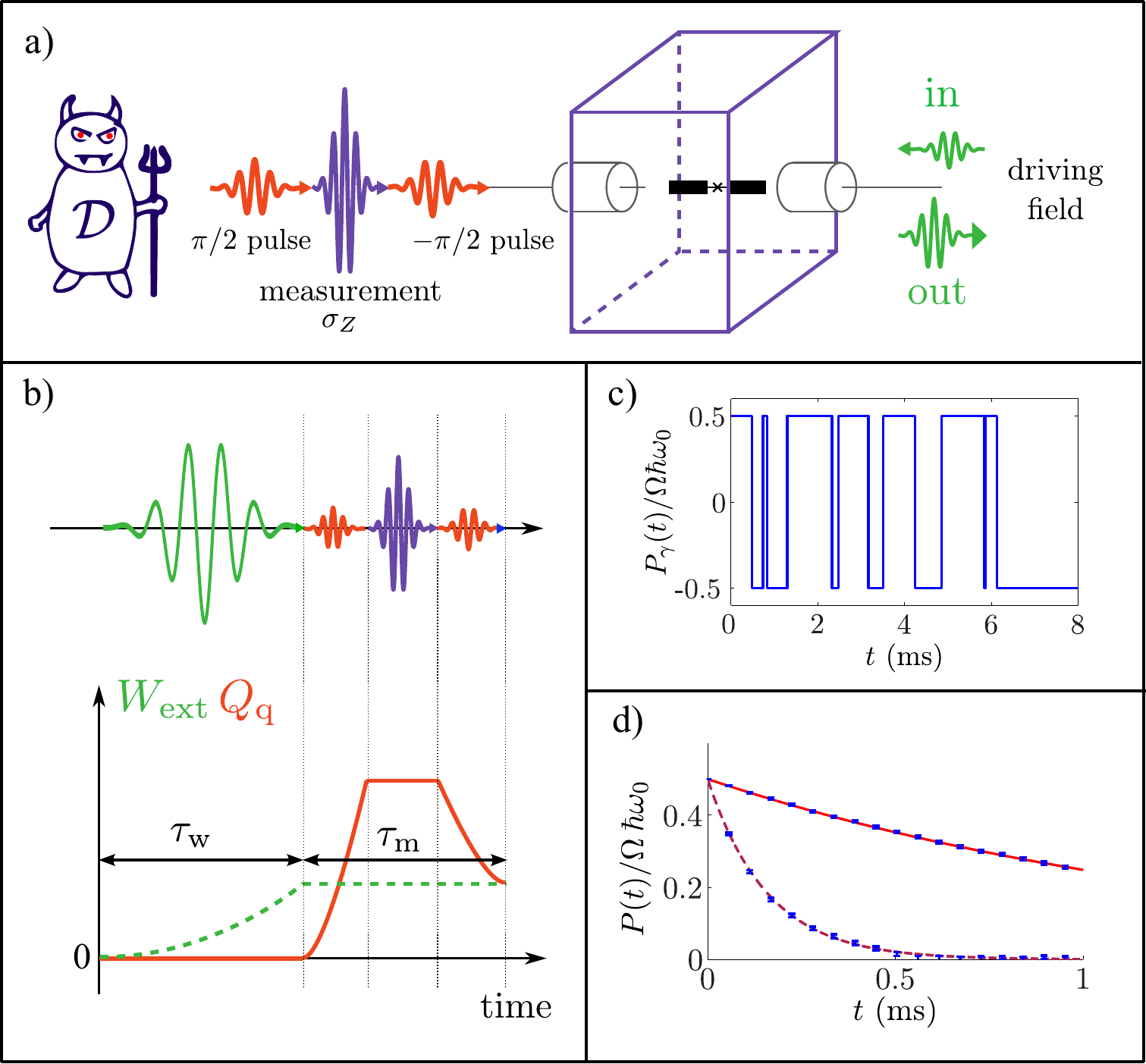}
\end{center}
\caption{Implementation of the MPE in circuit QED. a) Scheme of the experiment. The qubit is a transmon in a 3D cavity. Work is extracted as microwave propagating photons. b) Sequence of pulses corresponding to one engine cycle. Green: Drive. Orange: $\pi/2$-pulses. Purple: readout tone. Bottom: Work extracted (green dashed) and measurement energy (orange) as a function of time.  c) Extracted power $P_\gamma(t)$ in units of $\Omega \hbar\omega_0$ as a function of time for a single realization $\gamma$ of the engine. d) Mean power $P(t)$ as a function on time. The blue points are averaged over ${\cal N}_\text{r} = 10^4$ stochastic realizations. The size of the point corresponds to the error bar $\sigma_P(t)/\sqrt{{\cal N}_\text{r}}$, where $\sigma_P(t)$ is the variance of $P_\gamma(t)$. Red-solid and purple-dashed curves are obtained from the analytical formula provided in the Supplemental, for two different values of $\Omega$. {\it Parameters}: $\tau_\text{mes} = 70$ns, $\tau_\text{w} = 70$ns, $\Omega = 0.2$MHz (red solid) and $\Omega = 0.6$MHz (purple dashed).  \label{fig3}
}
\end{figure}

The work extraction step (i) is performed by driving the qubit with a microwave field through an input port of the cavity mode (Fig.~\ref{fig3}a). In the limit where the Rabi frequency $\Omega$ is much larger than the qubit decay rate $\gamma$, the extracted power is directly given by the difference between the output and input powers (See Ref.~\cite{Cottet-17} and Supplemental). As shown above, power extraction is maximized in the Zeno limit. Experimentally, this regime requires $\gamma^{-1} \gg \Omega ^{-1}\gg \tau_\text{w} \gg \tau_\text{mes}$: These conditions can be fulfilled with more than one order of magnitude between each timescale, as evidenced by the observations of the Zeno regime in circuit-QED setups \cite{Bretheau,Slichter}.

It is interesting to look at the implementation in the open loop protocol, i.e. in the absence of feedback. In Fig.~\ref{fig3}c we have plotted the simulated power $P_\gamma(t)$ extracted in a single stochastic realization $\gamma$ of the protocol, with realistic parameters taking into account the finite measurement time. One clearly sees the quantum jumps between the two working points of the engine, i.e. the state $\ket{+_\textbf{x}}$ (resp. $\ket{-_\textbf{x}}$) giving rise to a positive (resp. a negative) power extraction. Fig.~\ref{fig3}d features the mean extracted power $P(t) = \langle P_\gamma(t) \rangle_\gamma$ as a function of time, averaged over a large number of realizations for a qubit initialized in the $\ket{+_\textbf{x}}$ state. At short times, the engine provides work. At large times, the memory about the initial state is lost and the qubit ends up in a perfectly mixed state, such that the mean extracted power vanishes and the engine switches off. The smaller Rabi angle $\theta$, the later the switch off occurs. The analytical expression (see Supplemental) is plotted together with the numerical simulation. \\

Our modeling suggests a physical interpretation of the energy $E_\text{meas}$ provided by the measurement channel in a specific case. We analyze the first cycle of the engine in which the qubit starts in state $\ket{+_\textbf{x}}$.  After the work extraction step, the qubit ends up in $\ket{\psi(\tau_w)} = \cos(\theta/2)\ket{+_\textbf{x}}+\sin(\theta/2)\ket{-_\textbf{x}}$ of internal energy $(1-\sin(\theta))\hbar\omega_0/2$. The first $\pi/2$-pulse maps the qubit state onto $\ket{\psi^{'}} = \cos(\theta/2)\ket{1}+\sin(\theta/2)\ket{0}$ of internal energy $\cos^2(\theta/2)\hbar \omega_0$. The $\pi/2$ pulse therefore provides an amount of energy (work) $(\cos(\theta)+\sin(\theta))\hbar \omega_0/2$. The next operation is the measurement of $\sigma_Z$, which projects the qubit onto the state $\ket{1}$ (resp. state $\ket{0}$) with probability $\cos^2(\theta/2)$ (resp. $\sin^2(\theta/2)$), providing the energy $\sin^2(\theta/2)\hbar\omega_0$ (resp. $-\cos^2(\theta/2)\hbar\omega_0$). Finally, the second $\pi/2$ pulse brings the state $\ket{1}$ (resp. $\ket{0}$) back onto $\ket{+_\textbf{x}}$ (resp. $\ket{-_\textbf{x}}$), which costs the energy $-\hbar\omega_0/2$ (resp. $\hbar\omega_0/2$). As expected, the total energy provided during the effective readout in the case where $\ket{1}$ is found is $E_\text{meas} = \sin(\theta)\hbar\omega_0/2$. This analysis shows that this energy transfer actually represents a composite term gathering the deterministic contributions of the $\pi/2$ pulses (work) and of the stochastic contribution of the projective measurement of $\sigma_Z$. In the Zeno regime $\theta \to 0$, this composite term reduces to the exchanged energy  during the $\pi/2$ pulses. 

%In this limit, our engine is similar to an energy transducer, that deterministically converts energy from the measurement channel into work extracted in the driving field. This engine has no classical equivalent, 

This example illustrates the interesting questions raised by the thermodynamic nature of the energy transferred to a quantum system during its measurement. The denomination of quantum heat suggested by some of us \cite{Elouard-16} acknowledges the stochastic nature of the system dynamics during the measurement-induced energy changes. Decomposing the measurement process in different steps leads to splitting the energy provided by the measurement channel into work and heat contributions; however, such decomposition depends on the protocol, while the quantum heat will remain the same, as a byproduct of measurement postulate. The status of this postulate is still debated nowadays (See e.g. \cite{CSM,PT-Zurek,PT-AG}), giving rise to different approaches to build a consistent quantum thermodynamics. Eventually, these different perspectives reflect the various interpretations of quantum mechanics, which coexist without altering the efficiency of the theory. \\

\noindent {\it Conclusion}
We propose and study the performances of a genuinely quantum heat engine, where work is extracted from stochastic quantum fluctuations induced by the measurement process, instead of stochastic thermal fluctuations due to the coupling to a hot bath. We show that our engine is a versatile device whose behavior can be controlled with the measurement rate. In the Zeno regime of frequent measurements, the MPE behaves like an energy transducer providing efficiencies close to one and maximal powers. MPEs can readily be implemented in state of the art setups of circuit and cavity quantum electrodynamics, where measurement can be performed in arbitrary bases and the Zeno regime has already been evidenced \cite{Bretheau,Slichter}. \\

%Our findings are based on the analogy between a thermal bath and a measuring apparatus, which both behave as a source of randomness, energy and entropy. This analogy is fruitful as it allows analyzing the irreversibility that is induced by quantum measurement and quantum work extraction in MPE within the same framework. Reciprocally, the energetics of quantum measurement will bring new tools to investigate the sustainability of quantum processing in the presence of irreversible decoherence \cite{Elouard-16}.\\

\textbf{Acknowledgements} We thank Michel Brune, S\'ebastien J\'ezouin, Jean-Michel Raimond for enlightening discussions. This work was supported by the ANR under the grant 13-JCJC-INCAL and by the COST network MP1209 ``Thermodynamics in the quantum regime".

 \pagebreak
%\widetext
\onecolumngrid
\vspace{\columnsep}
\begin{center}
\textbf{\large Supplemental Material: Extracting work from quantum measurement in Maxwell demon engines}

\vspace{5mm}

\title{Extracting work from quantum measurement in Maxwell's demon engines}

\author{Cyril Elouard}
\affiliation{CNRS and Universit\'e Grenoble Alpes, Institut N\'eel, F-38042 Grenoble, France}

\author{David Herrera-Mart\'i}
\affiliation{CNRS and Universit\'e Grenoble Alpes, Institut N\'eel, F-38042 Grenoble, France}

\author{B. Huard}
\affiliation{Laboratoire de Physique, Ecole Normale Sup\'erieure de Lyon, 46 all\'ee d'Italie, 69364 Lyon Cedex 7, France}
\affiliation{Laboratoire Pierre Aigrain, Ecole Normale Sup\'erieure-PSL Research University, CNRS, Universit\'e Pierre et Marie Curie-Sorbonne Universit\'es, Universit\'e Paris Diderot-Sorbonne Paris Cit\'e, 24 rue Lhomond, 75231 Paris Cedex 05, France}

\author{Alexia Auff\`eves}
\email{alexia.auffeves@neel.cnrs.fr}
\affiliation{CNRS and Universit\'e Grenoble Alpes, Institut N\'eel, F-38042 Grenoble, France}

\end{center}

\vspace{\columnsep}
\twocolumngrid
%%%%%%%%%% Merge with supplemental materials %%%%%%%%%%
%%%%%%%%%% Prefix a "S" to all equations, figures, tables and reset the counter %%%%%%%%%%
\setcounter{equation}{0}
\setcounter{figure}{0}
\setcounter{table}{0}
\setcounter{page}{1}
\makeatletter
\renewcommand{\theequation}{S\arabic{equation}}
\renewcommand{\thefigure}{S\arabic{figure}}
\renewcommand{\bibnumfmt}[1]{[S#1]}
%\renewcommand{\citenumfont}[1]{S#1}
%%%%%%%%%% Prefix a "S" to all equations, figures, tables and reset the counter %%%%%%%%%%

\section*{I. Extracted power}

The work extracted during each cycle starting in state $\ket{+_\textbf{n}}$ reads:
\be
W_\text{ext} &=& \dfrac{\hbar\omega_0}{2}\big((1-\cos(\theta))\cos(\theta_\textbf{n})\nonumber\\
&&\quad\quad\quad+\sin(\theta)\sin(\theta_\textbf{n})\cos(\phi_\textbf{n})\big).
\ee

When $\ket{+_\textbf{n}}$ and $\ket{-_\textbf{n}}$ have different energies, the feedback step gives rise to some non zero work cost $W_\text{fb}$ that contributes (positively or negatively) to work extraction:
\be
W_\text{fb}=-\hbar\omega_0 \sin^2(\theta/2) \cos(\theta_\textbf{n})(1-\sin^2(\theta_\textbf{n})\sin^2(\phi_\textbf{n})).\nonumber\\
\ee

Finally, the average extracted power reads (neglecting the duration of the readout, feedback and erasure steps):

\be
P_\textbf{n}(\theta)&=&\dfrac{W_\text{ext}+W_\text{fb}}{\tau_w}\nonumber\\
&=& {\Omega}\dfrac{\hbar\omega_0}{2} \left(\dfrac{\sin(\theta)}{\theta}x_\textbf{n} + \dfrac{1-\cos(\theta)}{\theta}z_\textbf{n} y_\textbf{n}^2\right),
\ee
where we have introduced the Cartesian coordinates of the unit vector $\textbf{n}=(x_\textbf{n},y_\textbf{n},z_\textbf{n})$.

%%%The mean extracted power $P_{+_\textbf{x}}(\theta)$ reads
%%%\begin{equation}
%%%P_{+_\textbf{x}}(\theta) = \Omega\dfrac{\hbar\omega_0}2 \frac{\sin(\theta)}{\theta}.
%%%\end{equation}
%%%
%%%When $\ket{+_\textbf{n}}$ and $\ket{-_\textbf{n}}$ have different energies, the feedback step also gives rise to some energy exchange that contributes (positively or negatively) to work extraction. Taking into account this effect,

\section*{II. Open loop implementation}

We have computed an analytical expression of the mean extracted power at the N$^{th}$ cycle of the engine. We focus on the limit $\theta\ll 1$ such that we can restrict our analysis to second order in $\theta$. We assume that the engine has been initialized with the qubit in the pure state $\ket{+_\textbf{x}}$. After $N$ cycles without feedback the qubit is in the mixed state $\rho^{(N)}= p_+^{(N)}\ket{+_\textbf{x}}\bra{+_\textbf{x}}+p_-^{(N)}\ket{-_\textbf{x}}\bra{-_\textbf{x}}$, such that work extraction lead to a mean extracted work $\delta W^{(N)} = (p_+^{(N)}-p_-^{(N)})\theta \omega_0/2$. The state $\rho^{(N+1)}$ is obtained by applying unitary evolution $e^{-i\theta\sigma_Y/2}$ and removing off-diagonal terms in the basis $\{\ket{+_\textbf{x}},\ket{-_\textbf{x}}\}$. By using the recurrence relation  $p_+^{(N+1)}-p_-^{(N+1)} = \cos(\theta)(p_+^{(N)}-p_-^{(N)})$, we derive the mean extracted power at the N$^{th}$ step

\begin{eqnarray}
P^{(N)}= \frac{\delta W^{(N)}}{\tau_\text{w}+\tau_\text{mes}} = \frac{\Omega \hbar\omega_0}{2} \frac{\tau_\text{w}}{\tau_\text{w}+\tau_\text{mes}} \cos^N(\theta)\label{analytical}
\end{eqnarray}

\noindent The integrated mean work provided by the engine after $N$ cycles reads $\theta \hbar\omega_0 \frac{1-\cos(\theta)^N}{1- \cos(\theta)}$ and converges towards $W_\text{max} = \hbar\omega_0/\theta$. Naturally, this work diverges as $\theta \rightarrow 0$ as the engine never switches off in the Zeno regime.

\section*{III. Work extraction in circuit Quantum ElectroDynamics}
Here we recall some of the theoretical results exploited to model the experiment reported in \cite{Cottet-17}. A qubit is driven by a classical field of Rabi frequency $\Omega$ via the input port of a cavity. In case of large qubit-cavity detuning, the cavity mode can be adiabatically eliminated and the rate of input $ \langle b_\text{in}^\dagger b_\text{in} \rangle $ and output photons $\langle b_\text{out}^\dagger b_\text{out} \rangle $ verify

\begin{equation}
\langle b_\text{out}^\dagger b_\text{out} \rangle = \langle b_\text{in}^\dagger b_\text{in} \rangle + \sqrt{\gamma}( \langle b_\text{in}^\dagger \sigma_- \rangle + \langle \sigma_-^\dagger b_\text{in} \rangle) + \gamma \langle \sigma_-^\dagger \sigma_- \rangle,
\end{equation} 
\noindent where $\gamma$ stands for the spontaneous emission rate of the qubit in the output port, $ b_\text{in}$ (resp. $ b_\text{out}$) is the lowering operator in the input (resp. output) mode of the cavity. The input field is a coherent field $\ket{\beta_\text{in}}$, where $\beta_\text{in}$ is chosen real without loss of generality. In the classical limit $\beta_\text{in} \gg 1$, the third right-handed term can be neglected. The extracted power $P$ (in units of $\hbar \omega_0$) eventually reads 

\begin{equation}
\frac{P}{\hbar \omega_0}  = \langle b_\text{out}^\dagger b_\text{out} \rangle - \beta_\text{in}^2 =  \frac{\Omega}{2}\langle \sigma_X \rangle.
\end{equation} 

\noindent We have used that $ \sqrt{\gamma} \beta_\text{in} = \Omega/2$. As shown in the main text, the extracted power $P$ is maximal in the Zeno limit where the qubit is frozen in the state $\ket{+_\textbf{x}}$, where it equals $P/\hbar \omega_0 = \Omega/2$.

\section*{IV. Work extraction in Cavity Quantum Electrodynamics}
The drive is modeled by a coherent field $\ket{\alpha} = e^{-|\alpha|^2/2} \sum_n \frac{\alpha^n}{\sqrt{n!}} \ket{n} = \sum_n \alpha_n \ket{n}$ initially injected in the mode of a high quality factor cavity. We denote $\bar{n} = |\alpha|^2$ the mean number of photons and $\phi$ the phase of the coherent field, such that $\alpha = \sqrt{\bar{n}} e^{-i\phi}$. The coupling is described by the Jaynes-Cummings Hamiltonian $H = i(\hbar \Omega_0/2) (\sigma_- a^\dagger -a \sigma_-^\dagger )$, where $\Omega_0$ is the vacuum Rabi frequency. 

The situation studied in the main text is retrieved in the classical limit of large $\bar{n}$, such that this Hamiltonian becomes $H = (\hbar \Omega_0/2) \sqrt{\bar{n}} [\cos(\phi) \sigma_Y + \sin(\phi) \sigma_X]$, where $\sigma_X$ and $\sigma_Y$ are the Pauli matrices. The classical Rabi frequency reads $\Omega = \Omega_0 \sqrt{\bar{n}}$. We choose $\phi = 0$, such that the eigenstates are $\ket{+_\textbf{y}}$ and $\ket{-_\textbf{y}}$. We provide here a full quantum analysis valid for smaller numbers of photons. As we show below, work extraction translates into an increase of the mean photon number in the cavity mode. 

The calculations are similar to the pioneering theory of Gea-Banacloche \cite{Gea-90}: At the initial time the qubit and the mode are prepared in the state $\ket{+_\textbf{x},\alpha}$ where  $\ket{+_\textbf{x}}=(\ket{0}+\ket{1})/\sqrt{2}$. The initial state can be rewritten $\ket{\psi(0)} = \sum_n [\alpha_n \ket{1,n} +\alpha_n\ket{0,n}]/\sqrt{2}$, and evolves into the entangled atom-field state:

\begin{eqnarray}
\ket{\psi(t)}& =& \dfrac{1}{\sqrt{2}}\sum_{n=0}^\infty[ c_{n+1} \alpha_n \ket{1,n} + s_{n+1} \alpha_n \ket{0,n+1} \nonumber \\
&& + c_n \alpha_n \ket{0,n}] -  \dfrac{1}{\sqrt{2}}\sum_{n=1}^\infty s_n \alpha_n \ket{1,n-1}.
\end{eqnarray}

\noindent We have introduced $c_n = \cos(\Omega_0\sqrt{n} t/2)$ and $s_n = \sin(\Omega_0 \sqrt{n} t/2)$. Rewriting this state in the basis $\ket{+_\textbf{x}}$, $\ket{-_\textbf{x}}$, it yields

\begin{equation}
\ket{\psi(t)} = \ket{+_\textbf{x},\alpha_+} + \ket{-_\textbf{x},\alpha_-}
\end{equation}
\noindent with 
\begin{eqnarray}
\ket{\alpha_+} &=& \frac{1}{2}\sum_{n=0}^\infty \bigg( \alpha_n c_n+\alpha_n  c_{n+1}  - \alpha_{n+1}s_{n+1}\bigg)\ket{n}  \nonumber\\ 
&&+\sum_{n=1}^\infty \alpha_{n-1}s_{n}\ket{n},
\end{eqnarray}
and 
\begin{eqnarray}
\ket{\alpha_-} &=& \frac{1}{2}\sum_{n=0}^\infty \bigg( \alpha_n c_n-\alpha_n  c_{n+1}  + \alpha_{n+1}s_{n+1}\bigg)\ket{n} \nonumber\\
&&+ \sum_{n=1}^\infty \alpha_{n-1}s_{n}\ket{n}.
\end{eqnarray}\\

\noindent We now consider the limit $\theta = \Omega_0 \sqrt{\bar{n}} t\rightarrow 0$, %Acknowledging that in the classical limit the distribution of $\alpha_n$ is very peaked around $\bar{n}$,
such that we get at first order in $\theta$: $c_n =1$, $s_n = \Omega_0 \sqrt{n} t/2$. The two states of the cavity mode $\ket{\alpha_+}$ and $\ket{\alpha_-}$ respectively correlated to $\ket{+_\textbf{x}}$ and $\ket{-_\textbf{x}}$ read:
\begin{eqnarray}
\ket{\alpha_-} &=& \sum_{n=0}^\infty \alpha_n \bigg(\alpha\frac{\Omega_0t}{4}     + \frac{1}{\alpha}\frac{\Omega_0t}{4}n\bigg)\ket{n}\nonumber\\
&=& \dfrac{\theta}{4}\sum_{n=0}^\infty \alpha_n \bigg(1    + \frac{n}{\alpha^2}\bigg)\ket{n}\nonumber\\
\end{eqnarray}
and
\begin{eqnarray}
\ket{\alpha_+} &=& \sum_{n=0}^\infty \alpha_n\bigg(  1  - \alpha \frac{\Omega_0t}{4}  + \frac{1}{\alpha}\frac{\Omega_0t}{4}n\bigg)\ket{n}\nonumber\\
&=& \sum_{n=0}^\infty e^{-\vert\alpha\vert^2/2} \left(1-\frac{\theta}{4}\right)\bigg(  \dfrac{\alpha^n}{\sqrt{n!}} + \dfrac{\theta/(4\alpha)}{1-\theta/4}n  \dfrac{\alpha^{n-1}}{\sqrt{n!}}\bigg)\ket{n}\nonumber\\
&\simeq& \ket{\alpha+\theta/(4\alpha)},
\end{eqnarray}

\noindent where we have used that $\theta\ll 1$.  The states $\ket{\alpha_\pm}$ are not normalized: their norm encodes the probability to find respectively $[\pm_\textbf{x}]$ as a result of the $\sigma_\text{X}$ measurement. We can check that the probability to find $[+_\textbf{x}]$ goes to $1$ when $\theta\to 0$, and that the probability of finding $[-_\textbf{x}]$ fulfils $\langle \alpha_-\ket{\alpha_-} \simeq  \dfrac{\theta^2}{4}$, which agrees with the calculation provided in the main text. When the outcome $[+_\textbf{x}]$ is found, the new photon number in the field is $(\alpha+\theta/(4\alpha))^2$ which equals $\alpha^2 + \theta/2$  to first order in $\theta$. At each cycle, the cavity energy therefore increases by an amount $\hbar\omega_0\theta/2 = W_\text{ext}$.

\end{document}